%% file: paper.tex
\documentstyle[aps,prd,epsfig,theorem]{revtex}
\newcommand{\bee}{\begin{equation}}
\newcommand{\ee}{\end{equation}}
\newcommand{\beea}{\begin{eqnarray}}
\newcommand{\eea}{\end{eqnarray}}

\theoremstyle{break}    \newtheorem{The}{Theorem}
\theoremstyle{plain}    
\theoremstyle{plain}    \newtheorem{Rem}{Remark}
\theoremstyle{plain}    
{\theorembodyfont{\rmfamily}    \newtheorem{Pro}{Proof}[The] }
{\theorembodyfont{\rmfamily}    \newtheorem{Exa}{Example}[The] }

\preprint{SNUTP-02-026}
\begin{document}
%\draft
%\thispagestyle{empty}
%\def\thefootnote{\fnsymbol{footnote}}
%\setcounter{footnote}{1}

%\begin{flushright}
%SNUTP-02-026
%\end{flushright}

\title{ 
Perturbative improvement of staggered fermions using fat links
}
\author{Weonjong Lee}
\address{
  School of Physics,
  Seoul National University,
  Seoul, 151-747, South Korea
  }
%
%\author{Stephen Sharpe}
%
%\address{
%  Physics Department,
%  University of Washington,
%  Seattle, WA 98195-1560, USA 
%  }
%
\date{\today}
\maketitle
\begin{abstract}
  We study possibility of improving staggered fermions using various
  fat links in order to reduce perturbative corrections to the
  gauge-invariant staggered fermion operators.
  We prove five theorems on SU(3) projection, triviality in
  renormalization, multiple SU(3) projections, uniqueness and
  equivalence.
  As a result of these theorems, we show that, at one loop level, the
  renormalization of staggered fermion operators is identical between
  SU(3) projected Fat7 links and hypercubic links, as long as the
  action and operators are constructed by imposing the same
  perturbative improvement condition.
  In addition, we propose a new view of SU(3) projection as a tool of
  tadpole improvement for the staggered fermion doublers.
  As a conclusion, we present alternative choices of constructing fat
  links to improve the staggered fermion action and operators, which
  deserve further investigation.
\end{abstract}
%
%\pacs{11.15.Ha, 12.38.Gc, 12.38.Aw}
%
%
%\def\thefootnote{\arabic{footnote}}
%\setcounter{footnote}{0}
%\clearpage

\section{Introduction}
\label{sec:intr}
The staggered fermion formulation on the lattice has a number of
advantages for numerical studies.
In particular, it preserves part of the chiral symmetry, which is
essential to calculate the weak matrix elements for $\epsilon' /
\epsilon$ and to remove the unwanted additive renormalization of light
quark masses.
In addition, computational cost with staggered fermions is noticeably
cheaper than any other fermion formulation on the lattice.
The chiral symmetry insures that the discretization errors are
quadratic in lattice spacing $a$. 
By construction, the staggered fermions carry four degenerate 
flavors, which in itself is not a major problem.
The major problem is that the naive ({\em i.e.}~unimproved) staggered
fermion action allows flavor changing quark-gluon interactions, which
are relatively large \cite{ref:luo:0}.
These flavor changing interactions are quadratic in $a$ and 
get weaker as the lattice approaches the continuum.

Recently, in
\cite{ref:luo:0,ref:sinclair:0,ref:orginos:1,ref:lepage:0}, it was
understood that the dominant flavor changing interactions originate
from a gluon exchange between quarks when the gluon carries a high
transverse momentum.
This interaction is a pure lattice artifact of order $a^2$ and can be
removed by modifying the lattice action at tree level.
Numerical studies in \cite{ref:orginos:1} showed that the fat link
improvement at tree level reduces the flavor changing effects
noticeably.
In \cite{ref:lepage:0}, the concept of the fat link action (Fat7)
introduced in \cite{ref:orginos:1} was systematically re-interpreted in
terms of Symanzik improvement programme.
There are two kinds of ${\cal O}(a^2)$ error at the tree level: one is
the flavor changing interaction originated from one gluon exchange
between quarks \cite{ref:orginos:1} and the other is a flavor
conserving kinetic term.
Based on this observation, a proposal of how to correct both was made
in \cite{ref:lepage:0}.
In this paper, we adopt the same notation as in \cite{ref:lepage:0}
and we call the fat link introduced in
\cite{ref:orginos:1,ref:lepage:0} ``Fat7'' afterwards for notational
convenience.
Recently, the hypercubic (HYP) action (a new fat link action) was
introduced in \cite{ref:anna:0,ref:anna:1}.
The HYP fat link is constructed such that the HYP smearing smoothes
the gauge fields within the hypercubes attached to the original thin
link with SU(3) projection after each smearing.
The HYP fat link improvement reduces the flavor symmetry breaking
effects more efficiently than the Fat7 action.
A problem with staggered fermions is that the composite operators
receive relatively large perturbative corrections even at one loop.
In this paper, our improvement goal is to minimize the perturbative
corrections.
Recently, in Ref.~\cite{ref:golterman:0}, it was observed that the
contribution from the staggered fermion doublers (it was named
``doubler-tadpole'') to the renormalization is as large as that from
the usual gluon tadpoles introduced in \cite{ref:lepage:1}.
This implies that the fat links can also improve the perturbative
behavior efficiently, since the origin of doubler-tadpoles are the
same as that of the flavor changing interactions.
%%
%%Hence, in this paper, we study the various fat links from the
%%perspective of improving the perturbative behavior.
%%

In this paper, we prove that one-loop renormalization of the staggered
operators is identical between the SU(3) projected Fat7 links and the
HYP link.
We also propose alternative choices of fat links to improve the action
and operators for better perturbative behavior, which is relatively
cheap to implement in numerical simulations on the lattice.
This paper is organized as follows. 
In Sec.~\ref{sec:review}, we describe our notation for various fat
links and, briefly, review previous works.
Sec.~\ref{sec:wlee-theorem} is devoted to proof of five theorems on
the equivalence, uniqueness and triviality of the one-loop
renormalization for the HYP fat link and the SU(3) projected Fat7
links.
In Sec.~\ref{sec:interp}, we interpret the meaning of these theorems,
which leads to a proposal of how to construct fat links in order to
reduce the perturbative corrections most efficiently.
We close with some conclusions.

\section{Notations and review}
\label{sec:review}
First, we review the improvement of removing the flavor changing
interactions in Ref.~\cite{ref:lepage:0}.
We define a covariant second derivative as
\begin{eqnarray}
\Delta_\rho^{(2)} U_\mu(x) \equiv \frac{1}{u_0^2 a^2} \bigg(
        U_\rho(x) U_\mu(x + \hat{\rho}) U_\rho^\dagger (x + \hat{\mu} )
        - 2 u_0^2 U_\mu(x) +
        U_\rho^\dagger(x -\hat{\rho}) U_\mu(x -\hat{\rho}) 
        U_\rho (x - \hat{\rho} + \hat{\mu}) \bigg)
\end{eqnarray}
Using this definition, we define the smearing operator $L_\rho$
(prefactor) as
\begin{eqnarray}
L_\rho (\alpha) \cdot U_\mu \equiv \bigg( 1 + 
\alpha  \frac{ a^2 \Delta_\rho^{(2)} }{4}
\bigg) U_\mu
\end{eqnarray}
The smeared link, $L_\rho (\alpha=1) \cdot U_\mu$ is identical to the
thin link $U_\mu$ up to order $a^2$ but vanishes when a single
gluon carries a momentum $p_\rho = \pi/a$.
Using $L_\rho$, we can rewrite the Fat7 link as 
\begin{eqnarray}
V^L_\mu = \frac{1}{6} \sum_{ {\rm perm} (\nu,\rho,\lambda)}
        L_\nu (\alpha_1) \cdot \bigg( L_\rho ( \alpha_2 ) \cdot 
        \Big( L_\lambda (\alpha_3) \cdot U_\mu \Big) \bigg)
\label{eq:lepage:fat-link-1}
\end{eqnarray}
Here, $ {\rm perm} (\nu,\rho,\lambda) $ represents all the possible
permutations of $\nu,\rho,\lambda$ indices ($\nu \ne \rho \ne \lambda
\ne \mu$).
When $\alpha_i = 1$, the prefactors vanish at tree level when a single
gluon emission carries a momentum $ p_\nu = \pi/a $ for any $\nu \ne
\mu$.

Second, we review the flavor conserving improvement proposed in
Ref.~\cite{ref:lepage:0}.
We define a covariant first derivative as
\begin{eqnarray}
\Delta_\rho^{(1)} U_\mu(x) \equiv \frac{1}{2 u_0^2 a} \bigg(
        U_\rho(x) U_\mu(x + \hat{\rho}) U_\rho^\dagger (x + \hat{\mu} )
-       U_\rho^\dagger(x-\hat{\rho}) U_\mu(x -\hat{\rho}) 
        U_\rho (x - \hat{\rho} + \hat{\mu}) \bigg)
\end{eqnarray}
The ${\cal O}(a^2)$ corrections introduced when $U_\mu$ is replaced by
$V^L_\mu$ cancels all the tree level flavor changing interactions.
There is still a remaining ${\cal O}(a^2)$ error at low energy as a
result of introducing the $\Delta_\rho^{(2)}$'s of $V^L_\mu$.
This flavor conserving error can be removed by further modifying
$V^L_\mu$.
\begin{eqnarray}
V^L_\mu \rightarrow 
V^{L'}_\mu =  V^L_\mu - 
        \sum_{\rho \ne \mu} \frac{ a^2 (\Delta_\rho^{(1)})^2 }{4} U_\mu
\label{eq:lepage:fat-link-2}
\end{eqnarray}

Recently, Hasenfratz and Knechtli made an interesting proposal of
hypercubic blocking (HYP) in \cite{ref:anna:0,ref:anna:1}.
The basic form of smearing transformation is a SU(3) projected
modified APE blocking:
\begin{eqnarray}
\overline{V}^H_\mu &=&  \textbf{Proj}_{SU(3)} \big[ V^H_\mu \big]
\nonumber \\
V^H_\mu &= & \bigg[ (1 - \alpha'_1) U_\mu(x) + 
\nonumber \\
& &     \frac{\alpha'_1}{6} \sum_{\nu\ne\mu} \Big(
        \overline{M}_{\nu;\mu} (x) 
        \overline{M}_{\mu;\nu} (x + \hat{\nu}) 
        \overline{M}_{\nu;\mu}^\dagger (x + \hat{\mu} )
\nonumber \\
& &     + \overline{M}_{\nu;\mu}^\dagger (x-\hat{\nu}) 
        \overline{M}_{\mu;\nu} (x -\hat{\nu}) 
        \overline{M}_{\nu;\mu} (x - \hat{\nu} + \hat{\mu}) \Big) \bigg] 
\label{eq:HYP:fat-link-1}
\\
\overline{M}_{\mu;\nu} &=&  \textbf{Proj}_{SU(3)} \big[ M_{\mu;\nu} \big]
\nonumber \\
M_{\mu;\nu} &= & \bigg[ (1 - \alpha'_2) U_\mu(x) + 
\nonumber \\
& &     \frac{\alpha'_2}{4} \sum_{\rho\ne\mu,\nu} \Big(
        \overline{W}_{\rho;\nu,\mu} (x) 
        \overline{W}_{\mu;\rho,\nu} (x + \hat{\rho}) 
        \overline{W}_{\rho;\nu,\mu}^\dagger (x + \hat{\mu} )
\nonumber \\
& &     + \overline{W}_{\rho;\nu,\mu}^\dagger (x-\hat{\rho}) 
        \overline{W}_{\mu;\rho,\nu} (x - \hat{\rho}) 
        \overline{W}_{\rho;\nu,\mu} (x - \hat{\rho} + \hat{\mu}) \Big) \bigg] 
\label{eq:HYP:fat-link-2}
\\
\overline{W}_{\mu;\nu,\rho} &=&  \textbf{Proj}_{SU(3)} 
\big[ W_{\mu;\nu,\rho} \big]
\nonumber \\
W_{\mu;\nu,\rho} &= & \bigg[ (1 - \alpha'_3) U_\mu(x) + 
\nonumber \\
& &     \frac{\alpha'_3}{2} \sum_{\lambda\ne\mu,\nu,\rho} \Big(
        U_\lambda (x) 
        U_\mu (x + \hat{\rho}) 
        U_\lambda^\dagger (x + \hat{\mu} )
\nonumber \\
& &     + U_\lambda^\dagger (x-\hat{\rho}) 
        U_\mu (x - \hat{\rho}) 
        U_\lambda (x - \hat{\rho} + \hat{\mu}) \Big) \bigg] 
\label{eq:HYP:fat-link-3}
\end{eqnarray}
The hypercubic blocking is composed of three smearing transformations
with each accompanied by SU(3) projection.
By construction, the resulting fat link $\overline{V}_\mu^H$
mixes with thin links only from the hypercubes attached to the
original thin link $U_\mu$.
The corresponding gauge field $H_\mu$ is defined as
\begin{eqnarray}
\overline{V}^H_\mu(x) = \exp 
\big( i a H_\mu ( x + \frac{1}{2} \hat{\mu} ) \big)
\label{eq:HYP-gauge-field}
\end{eqnarray}
For more details, please refer to Refs.~\cite{ref:anna:0,ref:anna:1}.
\section{Theorems for the fat link improvement}
\label{sec:wlee-theorem}
Here we present a series of theorems to clarify the nature of the fat
links obtained using the smearing transformations given in
Eqs.~(\ref{eq:lepage:fat-link-1}), (\ref{eq:lepage:fat-link-2}) and
(\ref{eq:HYP:fat-link-1}).
%--------------------------
% General form of smearing.
%--------------------------
Let us consider a general form of the fat links.
The gauge fields are related to the thin links as
\begin{eqnarray}
U_\mu(x) = \exp \big( i a A_\mu (x+\frac{1}{2} \hat{\mu}) \big)
\end{eqnarray}
where the gauge coupling is absorbed into a redefinition of gauge
fields without loss of generality. We will set $a=1$ for notational
convenience.
We can write the general form of the fat links using gauge fields:
\begin{eqnarray}
V_\mu(x) &=& 1 + i \sum_{\nu,y} \Lambda^{(1)}_{\mu\nu} (x,y) A_\nu(y) +
        \sum_{\nu,\rho,y,z} \Lambda^{(2)}_{\mu\nu\rho} (x,y,z) 
        A_\nu(y) A_\rho(z)
        + {\cal O}(A^3)
\nonumber \\
&=&  1 + \Lambda^{(1)}_\mu \cdot A + 
\Lambda^{(2)}_\mu \cdot A^2 + {\cal O}(A^3)
\end{eqnarray}
Here, $ V_\mu $ represents the general form of the fat links such as
$V^L_\mu$, $V^{L'}_\mu$ and $V^H_\mu$.
We also introduce a notation for SU(3) projection:
\begin{eqnarray}
\overline{V}_\mu =  \textbf{Proj}_{SU(3)} \big[ V_\mu \big].
\label{eq:su(3)-proj}
\end{eqnarray}
The SU(3) projected link is related to the effective gauge field, $B_\mu$.
\begin{eqnarray}
\overline{V}_\mu(x) = \exp 
\big( i a B_\mu ( x + \frac{1}{2} \hat{\mu} ) \big)
\end{eqnarray}
Here, note that $B_\mu = \sum_a B_\mu^a T_a$.
We may express $B_\mu^a$ as a perturbative expansion in powers of
$A_\rho^a$ fields.
\begin{eqnarray}
  B_\mu^a &=& \sum_{n=1}^{\infty} B_\mu^{a(n)} 
  \nonumber \\
  &=& B_\mu^{a(1)} + B_\mu^{a(2)} + {\cal O}(A^3)
\end{eqnarray}
Here, $B_\mu^{(n)}$ represents a term of order $A^n$.

\subsection{SU(3) Projection}
\label{subsec:su(3)}
%-----------------------
% 0) su(3) projection.
%-----------------------
\begin{The}[SU(3) Projection${}^1$]
  \begin{enumerate}
  \item The linear term is invariant under SU(3) projection. 
    \begin{eqnarray}
      B_\mu^{(1)}(x) = \sum_{\nu,y} \Lambda^{(1)}_{\mu\nu} (x,y) A_\nu(y)
                   = \Lambda^{(1)}_{\mu} \cdot A
    \end{eqnarray}
  \item The quadratic term is antisymmetric in gauge fields.
    \begin{eqnarray}
      B_\mu^{c(2)} &=& \frac{1}{2} \sum_{a,b} f_{abc}
      \sum_{\nu,\rho} \sum_{y,z} \Lambda^{(2)}_{\mu\nu\rho} (x,y,z) 
      A_\nu^a(y) A_\rho^b(z)
      \nonumber \\
      B_\mu^{(2)} &=& -i \frac{1}{2} 
      \sum_{\nu,\rho,y,z} \Lambda^{(2)}_{\mu\nu\rho} (x,y,z) 
      \big[ A_\nu(y), A_\rho(z)]
    \end{eqnarray}
    where $f_{abc}$ are the antisymmetric SU(3) structure constants
    with non-zero values defined by $ [ T_a , T_b] = i f_{abc} T_c
    $.\footnote{ Theorems 1 and 2 were known to Patel and Sharpe and
    used in their perturbative calculations~\cite{ref:sharpe:0},
    although they did not present their derivation and
    details~\cite{ref:sharpe:1}. 
    Theorem 1 is also mentioned in~\cite{ref:bernard:0}, although
    their derivation and details are not
    presented~\cite{ref:degrand:0}. }
  \end{enumerate}
\label{theorem:su(3)}
\end{The}
\begin{Pro}
  For the SU(3) projection, we define $X$ as follows:
  \begin{eqnarray}
    X = {\rm Tr} \big( \overline{V}_\mu^\dagger V_\mu \big)
  \end{eqnarray}
  The SU(3) projection means that $ \overline{V}_\mu $ is determined
  such that it should maximize ${\rm Re}(X)$ and minimize $({\rm
    Im}(X))^2$ under the condition:
  \begin{eqnarray}
    {\rm sign} \big( \det( \overline{V}_\mu ) \big) =
      {\rm sign} \big( \det( V_\mu ) \big)
  \end{eqnarray}
  Using perturbation, we may expand $X$ in powers of gauge fields.
  \begin{eqnarray}
    X &=& \sum_{i=0}^{\infty} X^{(i)}
  \end{eqnarray}
  Here, $X^{(n)}$ is a term of order $A^n$.
  It is easy to show that $X^{(0)} = 3$ and $X^{(1)} = 0$.
  The $X^{(2)}$ term provides a condition to determine $B^{(1)}$.
  \begin{eqnarray}
    X^{(2)} = {\rm Tr} \Big[ -\frac{1}{2} \big( B^{(1)}_\mu - 
    \Lambda^{(1)}_\mu \cdot A \big)^2 
    + \frac{1}{2} \big(\Lambda^{(1)}_\mu \cdot A \big)^2 
    + \Lambda^{(2)}_\mu \cdot A^2 \Big]
  \end{eqnarray}
  Here, note that $X^{(2)}$ is real: $X^{(2)} = {\rm Re}(X^{(2)})$ and
  ${\rm Im}(X^{(2)})=0$.
  It is clear that $B^{(1)}_\mu$ should satisfy the following condition
  in order to maximize ${\rm Re}(X^{(2)})$.
  \begin{eqnarray}
    B^{(1)}_\mu = \Lambda^{(1)}_\mu \cdot A
  \end{eqnarray}
  This proves the first part of Theorem \ref{theorem:su(3)}.
\end{Pro}

\begin{Pro}
  The $X^{(3)}$ term is
  \begin{eqnarray}
    X^{(3)} &=& {\rm Tr} \bigg[ \frac{1}{2} 
    \Big\{ \big( \Lambda^{(1)}_\mu \cdot A - B^{(1)}_\mu \big), \ 
    B^{(2)}_\mu \Big\}
    - i \frac{1}{3} \big( B^{(1)}_\mu \big)^3
    + i \Lambda^{(3)}_\mu \cdot A^3 
    - i \frac{1}{2} \Big\{ B^{(1)}_\mu, \  
    \Lambda^{(2)}_\mu \cdot A^2 \Big\} \bigg]
    \nonumber \\
    &=& {\rm Tr} \bigg[ 
    - i \frac{1}{3} \big( \Lambda^{(1)}_\mu \cdot A \big)^3
    + i \Lambda^{(3)}_\mu \cdot A^3 
    - i \frac{1}{2} \Big\{ \Lambda^{(1)}_\mu \cdot A , \  
    \Lambda^{(2)}_\mu \cdot A^2 \Big\} \bigg]
  \end{eqnarray}
  The $X^{(3)}$ term may include an imaginary part.
  In other word, the leading contribution to the ${\rm Im}(X)$ could
  be of order $A^3$.
  The $X^{(3)}$ term, however, does not provide any clue to
  determine $B^{(2)}_\mu$ because its coefficient vanishes.
  Therefore, it is necessary to study the next higher order term,
  $X^{(4)}$:
  \begin{eqnarray}
    X^{(4)} &=& {\rm Tr} \bigg[ - \frac{1}{2} 
    \Big( B^{(2)}_\mu +
    i \frac{1}{2} \big( B^{(1)}_\mu \big)^2 +
    i \Lambda^{(2)}_\mu \cdot A^2 \Big)^2
    + Y  \bigg]
    \label{eq:X-(4)}
  \end{eqnarray}
  where $Y$ is defined as 
  \begin{eqnarray}
    Y &=& {\rm Tr} \bigg[ - \frac{1}{2}
    \Big( \frac{1}{2} \big( B^{(1)}_\mu \big)^2
    + \Lambda^{(2)}_\mu \cdot A^2 \Big)^2
    + \Lambda^{(4)}_\mu \cdot A^4
    + \frac{1}{2} \Big\{ B^{(1)}_\mu, \ 
    \Lambda^{(3)}_\mu \cdot A^3 \Big\}
    - \frac{1}{4} \Big\{ \big( B^{(1)}_\mu \big)^2, \ 
    \Lambda^{(2)}_\mu \cdot A^2 \Big\}
    - \frac{1}{8} \big( B^{(1)}_\mu \big)^4
    \bigg]
  \end{eqnarray}
  Here, note that $Y$ is a known constant term, because 
  $B^{(1)}_\mu$ is fixed by the $X^{(2)}$ term.
  The coefficient of the $B^{(3)}_\mu$ term vanishes when
  $B^{(1)}_\mu = \Lambda^{(1)}_\mu \cdot A$.
  The $B^{(4)}_\mu$ term vanishes because it is traceless.
  Therefore, it would be best if $B^{(2)}_\mu$ could satisfy
  the following condition:
  \begin{eqnarray}
    B^{(2)}_\mu =  -i \frac{1}{2} \big( B^{(1)}_\mu \big)^2 -
    i \Lambda^{(2)}_\mu \cdot A^2
    = J + K
	\label{eq:B(2)-cond}
  \end{eqnarray}
  However, Eq.~(\ref{eq:B(2)-cond}) can not be satisfied with 
	real $B^{a(2)}_\mu$ gauge fields.
  From the following relation,
  \begin{eqnarray}
    J &\equiv& -i \frac{1}{2} \big( B^{(1)}_\mu \big)^2 
    \nonumber \\
    &=& -i \frac{1}{4} \sum_{a,b} B^{a(1)}_\mu B^{b(1)}_\mu 
    \big( \frac{1}{3} \delta_{ab} + d_{abc} T_c \big)
  \end{eqnarray}
  note that the coefficient of the $J$ term in the adjoint representation
  is purely imaginary, whereas $B^{a(2)}_\mu$ is real.
  Hence, there is no way that $B^{(2)}_\mu$ can cancel off any part of
  the $J$ term.
  How about the $K$ term?
  \begin{eqnarray}
    K &\equiv& -i \Lambda^{(2)}_\mu \cdot A^2
    \nonumber \\
    &=& -i \sum_{a,b} \sum_{\nu,\rho} \sum_{y,z} 
	\Lambda^{(2)}_{\mu\nu\rho} (x,y,z) 
        A_\nu^a(y) A_\rho^b(z)
        \frac{1}{2} \Big( \frac{1}{3} \delta_{ab} + 
        (d_{abc} + i f_{abc}) T_c \Big)
  \end{eqnarray}
  The $K$ term contains a non-trivial real antisymmetric term in
  the adjoint representation.
  Thus, in order to maximize ${\rm Re}X^{(4)}$, $B^{(2)}_\mu$ must
  satisfy the following:
  \begin{eqnarray}
    B^{c (2)}_\mu = \frac{1}{2} \sum_{a,b} f_{abc}
        \sum_{\nu,\rho,y,z} \Lambda^{(2)}_{\mu\nu\rho} (x,y,z) 
        A_\nu^a(y) A_\rho^b(z)
  \end{eqnarray}
  This completes a proof of the second part of Theorem
  \ref{theorem:su(3)}.
\end{Pro}

\subsection{Renormalization at one loop}
\label{subsec:renorm}

\begin{The}[Triviality of one-loop renormalization]
  \begin{enumerate}
  \item At one loop level, only the $B_\mu^{(1)}$ term contributes
    to the renormalization of the gauge-invariant staggered fermion
    operators.
  \item At one loop level, the contribution from $ B_\mu^{(n)}$ for
    any $n \geq 2$ vanishes.
  \item At one loop level, the renormalization of the gauge-invariant
    staggered operators can be done by simply replacing the propagator
    of the $A_\mu$ field by that of the $B_\mu^{(1)}$ field.
  \end{enumerate}
  This theorem is true, regardless of details of the smearing
  transformation.
\label{theorem:renorm}
\end{The}

\begin{Exa}
  Let us consider gauge-invariant staggered bilinear operators
  as an example.
  The Feynman diagrams at one loop are given in
  Refs.~\cite{ref:sharpe:0,ref:jlqcd:0,ref:wlee:00}.
  Note that the $B_\mu^{(1)}$ contribution can be obtained by simply
  replacing the gauge field propagator of $\langle A_\mu(x) A_\nu(y)
  \rangle$ by $\langle B_\mu^{(1)}(x) B_\nu^{(1)}(y) \rangle$.
  The $B_\mu^{(2)}$ terms, in principle, may contribute to the tadpole
  diagrams (c) and (e) in Figure 1 of Ref.~\cite{ref:jlqcd:0}.
  However, this tadpole contribution from the $B_\mu^{(2)}$ term
  vanishes because the gauge field propagator $ \langle A_\nu^b(x)
  A_\rho^c(y) \rangle$ is symmetric in color indices (proportional to
  $\delta_{bc}$) and $f_{abc} \cdot \delta_{bc} = 0$.
  It is easy to show that the $B_\mu^{(n>2)}$ terms can not contribute
  to any one-loop diagrams.
  This proves Theorem \ref{theorem:renorm} for the gauge-invariant
  bilinear operators.
\end{Exa}

\begin{Pro}
  At one loop level, the $B_\mu^{(1)}$ terms can contribute to the
  renormalization of the gauge invariant operators exactly in the same
  way as the $A_\mu$ terms.
  In other words, the one-loop contribution from the $B_\mu^{(1)}$
  terms can be calculated by simply replacing the propagator of
  the $A_\mu$ field by that of the $B_\mu^{(1)}$ field.
  At one loop, the $B_\mu^{(2)}$ terms contributes only to the tadpole
  diagrams, whereas this is not true for the higher loop corrections.
  Here, the tadpole diagrams mean that the two $A_\mu$ gauge fields
  in the $B_\mu^{(2)}$ term are contracted with each other.
  The gauge field propagator $ \langle A_\nu^b(x) A_\rho^c(y) \rangle$
  is proportional to $\delta_{bc}$ and so symmetric with respect to
  the color index exchange: $b \leftrightarrow c$.
  However, the $B_\mu^{(2)}$ term is antisymmetric with respect to $b
  \leftrightarrow c$, whereas the gauge field propagator is symmetric.
  Therefore, the $B_\mu^{(2)}$ contribution to the one-loop
  renormalization vanishes, since $f_{abc} \cdot \delta_{bc} = 0$,
  regardless of details of the smearing transformations.
  The contribution from the $B_\mu^{(n>2)}$ terms to the one-loop
  renormalization vanishes because they are at least proportional to
\begin{eqnarray}
\langle A_\lambda \rangle^{n-2} = 
	\langle A_{\lambda_1}(x_1) \rangle
	\langle A_{\lambda_2}(x_2) \rangle
	\cdots
	\langle A_{\lambda_{n-2}}(x_{n-2}) \rangle
\end{eqnarray}
  and $\langle A_{\lambda_i} (x_i)\rangle = 0$ (the vacuum can not
  break Lorentz symmetry or rotational symmetry in QCD).
  Note that the information on the smearing transformation is
  contained only in $\Lambda_\mu^{(n)}$ and this proof is independent
  of $\Lambda_\mu^{(n)}$. In other words, this theorem is valid,
  regardless of details of the smearing transformation.
  This completes a proof of Theorem \ref{theorem:renorm}.
\end{Pro}

\subsection{Multiple SU(3) Projections}
\label{subsec:multi-su(3)}
As an example, let us consider $V^L_\mu$, the Fat7 link for
the flavor symmetry improvement.
One can perform a single SU(3) projection, or alternatively apply the
SU(3) projections in front of any $ L_\mu(\alpha_i)$ operators used to
construct $ V^L_\mu$, which we call ``multiple SU(3) projections''.
As in Eq.~(\ref{eq:su(3)-proj}), the single SU(3) projected Fat7 link
is defined as
\begin{eqnarray}
& & \overline{V}^L_\mu = \textbf{Proj}_{SU(3)} \big[ V^L_\mu \big] 
\nonumber \\ & &
\overline{V}^L_\mu(x) = \exp 
\big( i a C_\mu ( x + \frac{1}{2} \hat{\mu} ) \big)
\label{eq:single-su(3)-proj}
\end{eqnarray}
The multiple SU(3) projection is not unique. As an example, we may
define a multi-SU(3) projected Fat7 link as
\begin{eqnarray}
& & \overline{V}^M_\mu = \frac{1}{6} \sum_{perm(\nu,\rho,\lambda)}
        \textbf{Proj}_{SU(3)} \cdot L_\nu (\alpha_1) 
        \cdot \bigg( \textbf{Proj}_{SU(3)} \cdot L_\rho ( \alpha_2 ) \cdot 
        \Big( \textbf{Proj}_{SU(3)} \cdot L_\lambda (\alpha_3) \cdot 
        U_\mu \Big) \bigg)
\nonumber \\ & &
\overline{V}^M_\mu(x) = \exp 
\big( i a D_\mu ( x + \frac{1}{2} \hat{\mu} ) \big)
\label{eq:multi-su(3)-proj-1}
\end{eqnarray}
The other example is
\begin{eqnarray}
& & \overline{V}^{M'}_\mu = \frac{1}{6} \sum_{perm(\nu,\rho,\lambda)}
        \textbf{Proj}_{SU(3)} \cdot L_\nu (\alpha_1) 
        \cdot \bigg( \textbf{Proj}_{SU(3)} \cdot L_\rho ( \alpha_2 ) \cdot 
        \Big( L_\lambda (\alpha_3) \cdot 
        U_\mu \Big) \bigg)
\nonumber \\ & &
\overline{V}^{M'}_\mu(x) = \exp 
\big( i a D'_\mu ( x + \frac{1}{2} \hat{\mu} ) \big)
\label{eq:multi-su(3)-proj-2}
\end{eqnarray}
Another possibility is
\begin{eqnarray}
& & \overline{V}^{M''}_\mu = \frac{1}{6} \sum_{perm(\nu,\rho,\lambda)}
        \textbf{Proj}_{SU(3)} \cdot L_\nu (\alpha_1) 
        \cdot \bigg( L_\rho ( \alpha_2 ) \cdot 
        \Big( \textbf{Proj}_{SU(3)} \cdot L_\lambda (\alpha_3) \cdot 
        U_\mu \Big) \bigg)
\nonumber \\ & &
\overline{V}^{M''}_\mu(x) = \exp 
\big( i a D''_\mu ( x + \frac{1}{2} \hat{\mu} ) \big)
\label{eq:multi-su(3)-proj-3}
\end{eqnarray}

\begin{The}[Role of multiple SU(3) projections]
  \begin{enumerate}
  \item The linear gauge field term in the perturbative expansion is
    universal.
    \begin{eqnarray}
    C_\mu^{(1)} = D_\mu^{(1)} = {D'}_\mu^{(1)} = {D''}_\mu^{(1)}
    \end{eqnarray}
  \item In general, the quadratic terms may be different from one another.
    But all of them are antisymmetric in gauge fields.
    \begin{eqnarray}
      C_\mu^{(2)} &=& -i \frac{1}{2} 
      \sum_{\nu,\rho,y,z} \Omega_{\mu\nu\rho} (x,y,z) 
      \big[ A_\nu(y), A_\rho(z)]
      \nonumber \\
      D_\mu^{(2)} &=& -i \frac{1}{2} 
      \sum_{\nu,\rho,y,z} \Gamma_{\mu\nu\rho} (x,y,z) 
      \big[ A_\nu(y), A_\rho(z)]
      \nonumber \\
      {D'}_\mu^{(2)} &=& -i \frac{1}{2} 
      \sum_{\nu,\rho,y,z} \Gamma'_{\mu\nu\rho} (x,y,z) 
      \big[ A_\nu(y), A_\rho(z)]
      \nonumber \\
      {D''}_\mu^{(2)} &=& -i \frac{1}{2} 
      \sum_{\nu,\rho,y,z} \Gamma''_{\mu\nu\rho} (x,y,z) 
      \big[ A_\nu(y), A_\rho(z)]
    \end{eqnarray}
  \end{enumerate}
  This theorem is true, regardless of the details of the smearing
  operator $L_\mu(\alpha)$ and its parameter $\alpha$.
\label{theorem:multi-su(3)}
\end{The}

\begin{Pro}
  By Theorem \ref{theorem:su(3)}, the linear term of the smearing
  operators are the same under SU(3) projection.
  In other words, Theorem \ref{theorem:su(3)} tells us that the linear
  term of the $ L_\lambda(\alpha_3) $ operator is identical to that of
  the $ \textbf{Proj}_{SU(3)} \cdot L_\lambda(\alpha_3)$ operator.
  Then the repeated application of Theorem \ref{theorem:su(3)} tells
  us that the linear term of the $ L_\rho(\alpha_2) \cdot
  L_\lambda(\alpha_3) $ is identical to that of the following operators:
  \begin{enumerate}
  \item
    $L_\rho(\alpha_2) \cdot 
    \textbf{Proj}_{SU(3)} \cdot L_\lambda(\alpha_3)$
  \item
    $\textbf{Proj}_{SU(3)} \cdot L_\rho(\alpha_2) \cdot 
    L_\lambda(\alpha_3)$
  \item
    $\textbf{Proj}_{SU(3)} \cdot L_\rho(\alpha_2) \cdot 
    \textbf{Proj}_{SU(3)} \cdot L_\lambda(\alpha_3)$
  \end{enumerate}
  Applying Theorem \ref{theorem:su(3)} once more, we obtain the final
  results: the linear terms of the following 8 operators are the same.
  \begin{enumerate}
  \item
    $L_\nu(\alpha_1) \cdot 
    L_\rho(\alpha_2) \cdot 
    L_\lambda(\alpha_3)$
  \item
    $L_\nu(\alpha_1) \cdot 
    L_\rho(\alpha_2) \cdot 
    \textbf{Proj}_{SU(3)} \cdot L_\lambda(\alpha_3)$
  \item
    $L_\nu(\alpha_1) \cdot 
    \textbf{Proj}_{SU(3)} \cdot L_\rho(\alpha_2) \cdot 
    L_\lambda(\alpha_3)$
  \item
    $L_\nu(\alpha_1) \cdot 
    \textbf{Proj}_{SU(3)} \cdot L_\rho(\alpha_2) \cdot 
    \textbf{Proj}_{SU(3)} \cdot L_\lambda(\alpha_3)$
  \item
    $\textbf{Proj}_{SU(3)} \cdot L_\nu(\alpha_1) \cdot 
    L_\rho(\alpha_2) \cdot 
    L_\lambda(\alpha_3)$
  \item
    $\textbf{Proj}_{SU(3)} \cdot L_\nu(\alpha_1) \cdot 
    L_\rho(\alpha_2) \cdot 
    \textbf{Proj}_{SU(3)} \cdot L_\lambda(\alpha_3)$
  \item
    $\textbf{Proj}_{SU(3)} \cdot L_\nu(\alpha_1) \cdot 
    \textbf{Proj}_{SU(3)} \cdot L_\rho(\alpha_2) \cdot 
    L_\lambda(\alpha_3)$
  \item
    $\textbf{Proj}_{SU(3)} \cdot L_\nu(\alpha_1) \cdot 
    \textbf{Proj}_{SU(3)} \cdot L_\rho(\alpha_2) \cdot 
    \textbf{Proj}_{SU(3)} \cdot L_\lambda(\alpha_3)$
  \end{enumerate}
  This proves the first part of Theorem \ref{theorem:multi-su(3)}.
\end{Pro}

\begin{Pro}
  As long as the final operation is the SU(3) projection, we can apply
  Theorem \ref{theorem:su(3)}, which directly proves the second part
  of Theorem \ref{theorem:multi-su(3)}.
  Note that Theorem \ref{theorem:su(3)} is independent of 
  details of the smearing operator and so is this proof.
\end{Pro}

\subsection{Uniqueness of the Perturbative Improvement Program}
\label{subsec:uniqueness}
The perturbative improvement program for the flavor symmetry
restoration is a procedure of removing the flavor changing
interactions.
Both Fat7 and HYP programs are based upon the same philosophy of
maximally reducing the flavor changing interactions.
Since the technical details of how to construct the fat links are
different, they look different superficially.
Here, we address the question of whether the perturbative improvement
programs can be different from each other even though they share the
same underlying philosophy.
First, let us present the conclusion and explain the details later:
the answer is NO and the perturbative improvement program should be
unique at least at one loop level.
The following two theorems will prove this in two steps.

\begin{The}[Uniqueness]
  If we impose the perturbative improvement condition of
  removing the flavor changing interactions on the HYP action,
  the HYP gauge field, $H_\mu$ defined in Eq.~(\ref{eq:HYP-gauge-field}) 
  satisfies the following:
  \begin{enumerate}
  \item The linear gauge field term in perturbative expansion is
    identical to that of the SU(3) projected Fat7 links.
    \begin{eqnarray}
    H_\mu^{(1)} = C_\mu^{(1)} = D_\mu^{(1)} = {D'}_\mu^{(1)} = {D''}_\mu^{(1)}
    \label{eq:unique-1}
    \end{eqnarray}
    where $C_\mu$, $D_\mu$, ${D'}_\mu$ and ${D''}_\mu$ are defined
    in Eqs.~(\ref{eq:single-su(3)-proj}-\ref{eq:multi-su(3)-proj-3}).
  \item The quadratic terms are antisymmetric in gauge fields.
    \begin{eqnarray}
      H_\mu^{(2)} & = & -i \frac{1}{2} 
      \sum_{\nu,\rho,y,z} \Xi_{\mu\nu\rho} (x,y,z) 
      \big[ A_\nu(y), A_\rho(z)]
    \end{eqnarray}
  \end{enumerate}
\label{theorem:unique}
\end{The}

\begin{Pro}
  The perturbative improvement condition of removing the flavor
  changing interactions imposes the three restrictions: one gluon
  emission vertex vanishes at the momentum $q_\nu = \pi/a$ for
  $\forall \ \nu \ne \mu$ (any direction transverse to the
  original link direction, $\mu$).
  These three restrictions fix the coefficients of Fat7 links as
  $\alpha_1 = \alpha_2 = \alpha_3 = 1$ at tree level, regardless of
  the SU(3) projection.\footnote{Note that $u_0=1$ at tree level.}
  The same restrictions determine the coefficients of the HYP
  blocking: $\alpha'_1 = 7/8$, $\alpha'_2=4/7$ and $\alpha'_3 = 1/4$.
  The key point is that the improvement condition is necessary and
  sufficient to determine all the coefficients so that the linear
  terms satisfy the universal relation given in Eq.~(\ref{eq:unique-1}).
  One may ask why there is no ambiguity in the flavor conserving
  terms.
  By construction, the HYP link incorporates the gauge degrees of 
  freedom within the hypercubes attached to the original link.
  As in Eq.~(\ref{eq:lepage:fat-link-2}), the flavor conserving
  kinetic terms, however, goes beyond the hypercubes attached to the
  original link.
  Hence, there was no ambiguity originated from the flavor conserving
  interactions.
  This proves the first part of Theorem \ref{theorem:unique}.
\end{Pro}

\begin{Pro}
  By construction, the HYP fat links are SU(3)-projected after each
  smearing transformation as shown in
  Eq.~(\ref{eq:HYP:fat-link-1}--\ref{eq:HYP:fat-link-3}).
  By Theorem \ref{theorem:su(3)}, the quadratic terms of any SU(3)
  projected fat links are antisymmetric in gauge fields.
  This proves the second part of Theorem \ref{theorem:unique}.
\end{Pro}

\begin{The}[Equivalence at one loop]
  If we impose the perturbative improvement condition to remove the
  flavor changing interactions, at one loop level,
  \begin{enumerate}
  \item the renormalization of the gauge invariant staggered operators
    is identical between the HYP staggered action and those improved
    staggered actions made of the SU(3) projected Fat7 links,
  \item and the contribution to the one-loop renormalization can be
    obtained by simply replacing the propagator of $A_\mu$ field by
    that of the $H_\mu^{(1)}= C_\mu^{(1)} = D_\mu^{(1)} =
    {D'}_\mu^{(1)}= {D''}_\mu^{(1)}$ field.
  \end{enumerate}
  Here, the SU(3) projected Fat7 links collectively represent the
  $\overline{V}^L_\mu$, $\overline{V}^M_\mu$, $\overline{V}^{M'}_\mu$
  and $\overline{V}^{M''}_\mu$ gauge links defined in
  Eqs.~(\ref{eq:single-su(3)-proj}-\ref{eq:multi-su(3)-proj-3}).
  \label{theorem:equivalence}
\end{The}

\begin{Pro}
  The quadratic terms in the HYP fat links and the SU(3) projected
  Fat7 links make no contribution to one loop renormalization by
  Theorem \ref{theorem:renorm}.
  Since the quadratic and higher order terms can not contribute, only
  the linear terms contribute to the renormalization at one loop by
  Theorem \ref{theorem:renorm}.
  The linear terms are identical between the HYP fat links and the
  SU(3) projected Fat7 links by Theorem \ref{theorem:unique}.
  Therefore, the one-loop renormalization of the gauge invariant
  staggered fermion operators are the same between the HYP and SU(3)
  projected Fat7 links.
  Since only the linear terms contribute, the renormalization
  constants can be calculated by simple replacement of the gauge
  propagator, $\langle A_\mu(x) A_\nu(y) \rangle$ with $\langle
  H_\mu^{(1)}(x) H_\nu^{(1)}(y) \rangle$ by Theorem
  \ref{theorem:renorm}.
  This completes a proof of Theorem \ref{theorem:equivalence}.
\end{Pro}

\section{Interpretation of the Theorems}
\label{sec:interp}
In the original definition of the Fat7 links, the $SU(3)$ projection
was not included in Ref.~\cite{ref:lepage:0}, because the improvement
aims at removing the ${\cal O}(a^2)$ terms based upon the Symanzik
improvement programme.
Hence, the flavor symmetry restoration was only a part of the
improvement goal in Ref.~\cite{ref:lepage:0}.
However, in this paper we have a somewhat different goal to minimize
the perturbative corrections to staggered fermion operators by
constructing the action and operators using fat links.

It has been a long-standing problem that the naive ({\em
i.e.}~unimproved) staggered fermion operators receive a large
perturbative corrections.
Recently, in \cite{ref:golterman:0}, it was pointed out that the
conventional tadpole improvement program suggested in
\cite{ref:lepage:1} works well for the Wilson fermions but it does not
work for the staggered fermions.
It was observed that the contribution from the staggered fermion
doublers (it is called ``doubler-tadpole'' in \cite{ref:golterman:0})
are very much like the usual gluon tadpoles originally introduced in
\cite{ref:lepage:1}.
Therefore, subtracting only the usual gluon tadpoles was not enough 
to improve the perturbative behavior.
Note that the origin of the doubler-tadpoles are the same as
that of the flavor changing interactions at the tree level.
This observation of the doubler-tadpole problem guided us to an
improvement program of removing the doubler-tadpoles systematically,
which is identical to the idea of removing the flavor changing
quark-gluon vertex suggested in
\cite{ref:orginos:1,ref:lepage:0,ref:orginos:0}.
Therefore, in this section, we will focus on the improvement programs
for the flavor symmetry restoration: the Fat7 improvement and the HYP
improvement, and interpret the meaning of the theorems given in
Sec. \ref{sec:wlee-theorem}.
\subsection{SU(3) Projection and Renormalization}
\label{subsec:su(3)-renorm}
Theorems \ref{theorem:su(3)} and \ref{theorem:renorm} tell us that the
SU(3) projection makes the one-loop renormalization of the staggered
operators so simple that it can be calculated simply by substituting
the fat link gauge propagator $\langle B_\mu^{(1)}(x) B_\nu^{(1)}(y)
\rangle$ for the original gauge propagator $\langle A_\mu(x) A_\nu(y)
\rangle$ in calculating each Feynman diagram.
By construction, the smearing prefactor of the fat link is designed to
suppress the high momentum gluon exchange.
Therefore, the SU(3) projected fat links guarantees that the
perturbative correction will be smaller than that of the original 
thin links.
\begin{Rem}[Tadpole improvement by the SU(3) projected fat links]
  Let us define $C_{fat}$ as the perturbative correction to the
  gauge invariant staggered fermion operators constructed using 
  the SU(3) projected fat links (Fat7 type and HYP type).
  Similarly, $C_{thin}$ is defined as the perturbative correction to
  the gauge invariant staggered fermion operator constructed using the
  usual thin links.
  Then, at one loop level, for each Feynman diagram,
  \begin{eqnarray}
    \parallel C_{fat} \parallel \ < \ \parallel C_{thin} \parallel
  \end{eqnarray}
\label{remark:tadpole}
\end{Rem}
Here, note that this remark does not apply to the Fat7 link without
SU(3) projection.\footnote{ Explicit calculation of one-loop diagrams
shows that the correction from the tadpole diagram is the same between
the thin link and Fat7 link without SU(3) projection even after the
first level of the tadpole improvement \cite{ref:wlee:0}. With no
tadpole improvement, the correction using the Fat7 link was even
larger.}
This is a direct consequence of Theorem \ref{theorem:su(3)} and
Theorem \ref{theorem:renorm} combined with the assumption that the
smearing prefactor suppresses the high momentum gluon exchange which
causes the doubler-tadpole problem.
Therefore, we conclude that the SU(3) projection of the fat links
(Fat7 type and HYP type) consistently decrease the contribution from
the doubler tadpoles by the ratio of $\parallel B_\mu^{(1)} / A_\mu
\parallel^2$.
We may view the SU(3) projection of the fat links as one way of
performing the tadpole improvement to remove the doubler tadpoles.
However, note that Remark \ref{remark:tadpole} does not guarantee that
the total summation of the one-loop corrections is smaller with the
SU(3) projected fat links, because the smallness of individual terms
does not mean much to the case of destructive cancellation between
Feynman diagrams.

\subsection{Uniqueness and Equivalence}
\label{subsec:unique-equiv}
From Theorems \ref{theorem:multi-su(3)}, and
\ref{theorem:equivalence}, we learn that the fat links with a single
SU(3) projection and multiple SU(3) projections make no difference to
one-loop renormalization of the gauge-invariant staggered fermion
operators.
For example, staggered operators made of the various fat links such as
$\overline{V}^L_\mu$, $\overline{V}^M_\mu$, $\overline{V}^{M'}_\mu$,
$\overline{V}^{M''}_\mu$ receive the same one-loop renormalization.
In addition, Theorems \ref{theorem:unique} and
\ref{theorem:equivalence} tells us that as long as we impose the same
perturbative improvement condition, the HYP staggered action and the
improved staggered actions of the SU(3) projected Fat7 type give the
same one-loop correction to the staggered operators, although the HYP
fat links are constructed in such a way completely different from the
SU(3) projected Fat7 links.

As a result of Theorems
\ref{theorem:su(3)}--\ref{theorem:equivalence}, at one loop level, we
have five equivalent choices for the improvement using fat links: the
four SU(3) projected Fat7 links and the HYP link.
All of them share the same advantages of triviality in
renormalization, smaller one-loop corrections for each Feynman
diagram, uniqueness and equivalence, which are significantly better
properties that the original Fat7 link action without SU(3) projection
does not possess.
However, one may still ask which of these five fat links are the best.
We will address this question next.

\section{The final proposal and conclusion}
\label{sec:final-proposal}
We know from Theorems \ref{theorem:su(3)}-\ref{theorem:equivalence}
that one-loop perturbation can not make any distinction between the
SU(3) projected Fat7 links and the HYP link.
In other words, one-loop perturbation can not guide us any more.
Let us first make a list of advantages of various fat links of our
concern.
The single SU(3) projected Fat7 link, $\overline{V}^L_\mu$ defined in
Eq.~(\ref{eq:single-su(3)-proj}) is relatively cheap to calculate on
the computer compared with the other fat links.
In Ref.~\cite{ref:anna:0}, it is shown that the $N=3$ APE smearing
\cite{ref:ape:0} and HYP blocking reduce the flavor symmetry breaking
in the pion spectrum more efficiently than the $N=1$ APE smearing,
which is consistent with Ref.~\cite{ref:orginos:1}.
This numerical results lead us to the conclusion that the fat link,
$\overline{V}^M_\mu$ defined in Eq.~(\ref{eq:multi-su(3)-proj-1}) and
the HYP fat link would probably be the best from the perspective of
the flavor symmetry restoration.
The fat link, $\overline{V}^M_\mu$ has an advantage that the
interpretation of the improvement is relatively straight-forward,
compared with the HYP fat link.
In addition, the $\overline{V}^M_\mu$ link is computationally simpler
to program, whereas the HYP link needs considerably more memory in
order to make it fast enough.\footnote{In order to make the HYP
blocking run fast, one needs to precompute the second level and the
third level of the HYP blocking and save them in on-board
memory~\cite{ref:anna:2}.  This requires considerably more memory than
the case of the triple SU(3) projected Fat7 links.}
%
%%%%%%%%%%%%%%%%%%%%%%%%%%%%%%%%%%%%%%%%%%%%%%%%%%%%%%%%%%%%%%%%%%%%
Considering all the advantages mentioned above, we make two final
proposals.
\begin{itemize}
\item Use the single SU(3) projected Fat7 link $\overline{V}^L_\mu$,
if the goal of improvement is to achieve, in a numerically cheaper
way, the smaller one-loop correction with all the nice features
mentioned in the previous section \ref{sec:interp}.
\item Use the triple SU(3) projected Fat7 link $\overline{V}^M_\mu$,
if the goal of improvement is to achieve the smaller one-loop
correction and the better flavor symmetry restoration simultaneously
with all the nice features.
\end{itemize}
%%%%%%%%%%%%%%%%%%%%%%%%%%%%%%%%%%%%%%%%%%%%%%%%%%%%%%%%%%%%%%%%%%%%
%
These two alternatives certainly deserve further investigation.

In addition, the five theorems in this paper make the perturbative
calculation simpler for the HYP type mainly because one can perform
the calculation merely by replacing the thin link propagator with that
of the HYP fat link.
This simplicity is extensively used in calculating one loop
renormalization constants of staggered fermion operators made of the
HYP links \cite{ref:wlee:0,ref:wlee:1}.

%
% EDIT
%

\section*{Acknowledgments}
\label{sec:acknowledge}
We would like to thank the INT (University of Washington, Seattle) and
its program organizers (S.~Sharpe and M.~Golterman), where this work
was begun.
% 
% We would like to express a sincere gratitude to S. Sharpe and
% M. Golterman for the invitation to the INT workshop held in University
% of Washington, Seattle with a title of ``Lattice QCD and Hadron
% Phenomenology'' (September 24 -- December 7, 2002).
%
We would like to express our heartfelt gratitude to S.~Sharpe for
several helpful discussions and for reading the manuscript.
We are grateful to T.~DeGrand and A.~Hasenfratz for helpful comments.
%
% During this workshop, WL picked up a premature form of the original
% idea of this paper through a series of a helpful discussion with
% S. Sharpe on the renormalization of the bilinear operators using
% various improved staggered fermions \cite{ref:wlee:0}, which was one
% of the prerequisite steps for the staggered $\epsilon'/\epsilon$
% project.
%
We would like to thank T.~Bhattacharya, N.~Christ, R.~Gupta, G.~Kilcup
and R.~Mawhinney for their support on the staggered $ \epsilon' /
\epsilon $ project. This work was supported in part by the BK21
program at Seoul National University, by the SNU foundation \&
Overhead Research fund and by Korea Research Foundation (KRF) through
grant KRF-2002-003-C00033.
%
%
%

\input{ref.tex}

\end{document}

%% file: ref.tex
%%%%%%%%%%%%%%%%%%%%%%%%%%%%%%%%%%%%%%%%%%%%%%%%%%%%%%%%%%%%%%%%%%%%%%
%  References
%%%%%%%%%%%%%%%%%%%%%%%%%%%%%%%%%%%%%%%%%%%%%%%%%%%%%%%%%%%%%%%%%%%%%%
%\begin{thebibliography}{99}